\documentclass[pre,amsmath,twocolumn, amssymb,notitlepage]{revtex4}

\usepackage{dcolumn}
\usepackage{lipsum}
\usepackage{graphicx}
\usepackage{placeins}
\usepackage{float}
\usepackage{bm}
\usepackage[T1]{fontenc}
\usepackage{color}
\usepackage{url}
\usepackage[bf]{subfigure}
\usepackage{rotating}
\usepackage{amsfonts}
\usepackage{multirow}	
\usepackage{xcolor}
\usepackage[utf8]{inputenc}

\begin{document}

\title{Note: Distance-Based Network Partitioning}

\author{Paulo J. P. de Souza }
\author{Cesar H. Comin}
\author{Luciano da F. Costa}

\affiliation{$^1$Instituto de Física de São Carlos, Universidade de São Paulo, São Carlos, São Paulo, Brazil\\
$^2$Universidade Federal de São Carlos, São Carlos, São Paulo, Brazil}

\begin{abstract}
A new method for identifying soft communities in networks is proposed. Reference nodes, either selected using a priori information about the network or according to relevant node measurements, are obtained. Distance vectors between each network node and the reference nodes are then used for defining a multidimensional coordinate system representing the community structure of the network at many different scales. For modular networks, the distribution of nodes in this space often results in a well-separated clustered structure, with each cluster corresponding to a community.  The potential of the method is illustrated with respect to a spatial network model and the Zachary's karate club network.
\end{abstract}

\maketitle

\section{Introduction}

One of the interesting properties of several real-world complex networks --- such as scientific collaborations, brain networks, social and economical networks --- concerns their modular structure.  Modularity is important from both topological and dynamical points of view~\cite{fortunato2010community}.  Topologically, communities correspond to the partitioning of the network into major groups of reference, revealing much about the possible origin of the communities as well as the behavior of different dynamics in the network.  Indeed, modularity influences dynamics, because it tends to constrain dissemination of activation inside each module (e.g.~\cite{costa2009beyond}).  It was also shown that the all important issue regarding the interplay between topology and dynamics is heavily influenced by modularity, in the sense that different types of such a relationship can be observed within communities of a \emph{same} network~\cite{comin2014random}.  

For all its importance and promises, modularity remains a challenge as a consequence of the difficulty of, given a network, to identify its respective communities~\cite{fortunato2010community}.  Indeed, many are the reported approaches proposing new methods of community detection (e.g.~\cite{fortunato2010community}).  Part of the difficulty in finding communities are better understood by taking into account the direct analogy between this task and the problem of \emph{clustering}, or \emph{unsupervised classification}, in the research area known as pattern recognition, which is backed by decades of investigations (e.g.~\cite{nasrabadi2007pattern}).  The main problem in clustering concerns the diversity of manners in which a cluster can appear or be defined.  Basically, a cluster is a subset of the objects so that its elements are more similar (closer) one another than with objects in the remainder of the set.  The multiplicity of ways to define a cluster can start to be appreciated by observing that there is an infinite number of possible distance types in a geometric space (e.g. Euclidean, Mahalanobis, cityblock, etc.).  Distinct metrics will possibly be more compatible with specific types of data.  

Another problem in defining clustering is the criteria to be adopted for deciding if each of the given objects belong or not to that cluster.  This issue can be effectively treated by mapping the objects into a density space, so that clusters become associated to density extremes. However, as a consequence of real-world complexity, these extremes will appear along a whole set of spatial scales.  These two main types of problems, choice of metrics and cluster identification, are directly inherited by community detection approaches.  

The current work addresses the problem of community detection through an analogy with prototype-based supervised classification.  More specifically, a prototype node is assigned to each community, and used as a reference from which respective distances are calculated.

\section{Materials and methods}
\subsection{Our method}

Our method requires the definition of references nodes. These central nodes can be defined using a priori information about the network or according to measurements such as the degree, clustering coefficient, accessibility~\cite{costa2007characterization}, and/or geographic characteristics of the network. Yet another possibility is to have these reference nodes provided by the user. Given the reference nodes, a distance vector $D_i$ is defined for each node $i$ in the network. The $j$-the element of $D_i$ contains all the distances between node $i$ and the reference node $c_j$ times a weight factor $\alpha_{c_j}$. That is

\begin{equation}
\vec{D}_i = (\alpha_{c_1}d_{ic_1}, \alpha_{c_2}d_{ic_2},...,\alpha_{c_M}d_{ic_M}),
\end{equation}
where $M$ is the number of reference nodes. Distance $d_{ic_j}$ can be topological or spatial. 

Each of the reference nodes can be understood as a kind of coordinate axis in a multidimensional space  $\Omega$.  The consideration of the several distances, taking into account their multiplicity, provides robust indication about the position of each of the network nodes with respect to the overall network structure, typically inducing respective well-separated clusters in $\Omega$.  Each of these clusters tend to correspond to one of the original communities.  The assignment of each node to each of the communities is done by taking into account the minimum topological or spatial distance with each of the reference nodes. Ties were broken by comparing the sum of distances of the neighbors of the node to each center.

\subsection{Benchmark networks}
\label{s:benchmark}

A benchmark network model was defined in order to verify if the method proposed in this work can correctly identify the community structure of spatial networks. The model works as follows: first, a set of reference points $S$ is defined. For each reference point $p$, a Poisson point process~\cite{chiu2013stochastic} with density $\lambda_p$ is performed inside a circle of radius $R_p$ centered on the point, defining the positions of the network nodes. Nodes are then connected according to the Waxman~\cite{WX} algorithm, that is, node $i$ is connected to node $j$ with probability

\begin{equation}
p_{ij} = e^{-\beta d_{ij}}
\end{equation}
where $d_{ij}$ is the Euclidean distance between the nodes and $\beta$ is a constant that controls the network average degree.

\section{Preliminary Results}

\begin{figure*}[ht]
	\centering
  	\subfigure[]{
    \includegraphics[width=0.7\columnwidth]{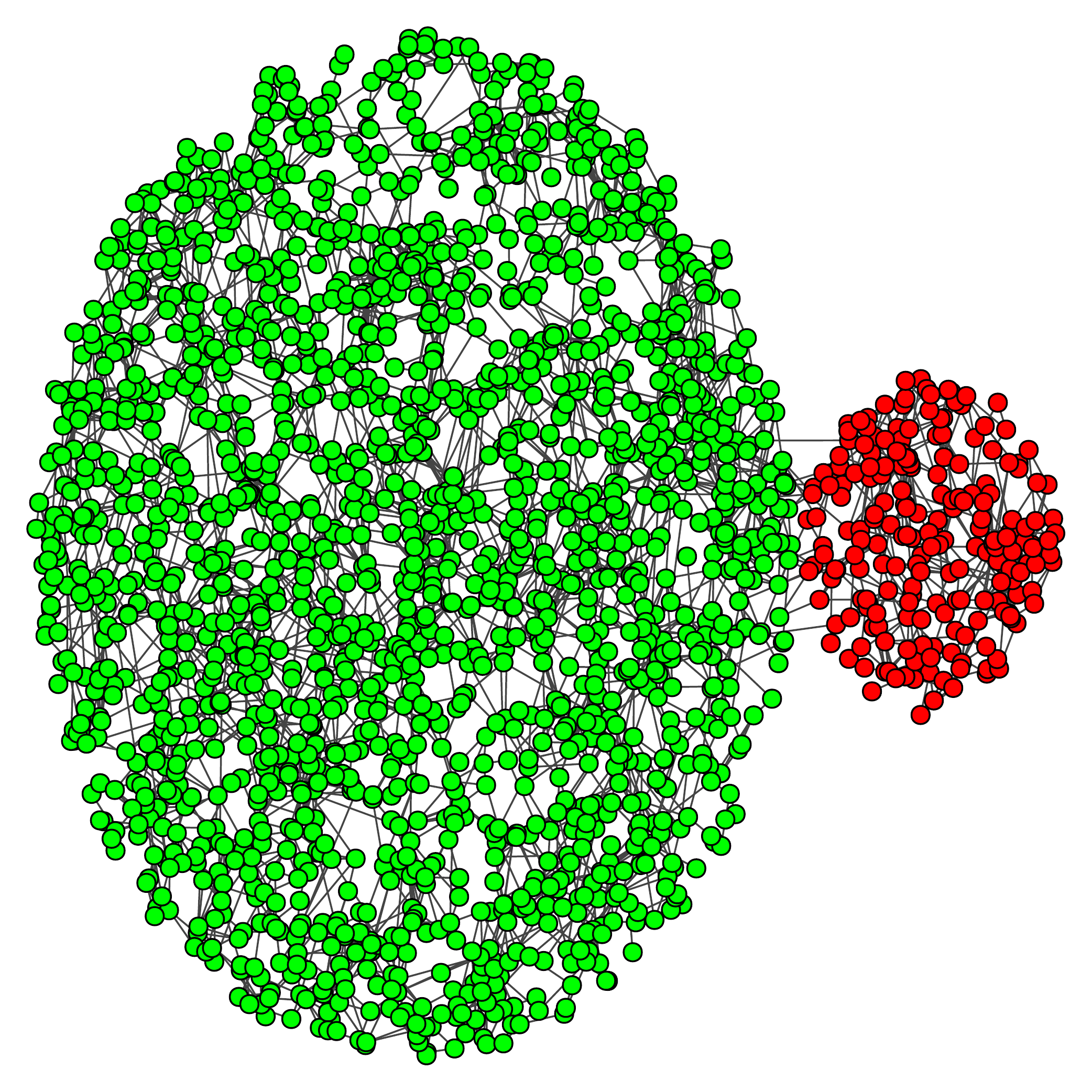} 
    \label{fig:2_wx_a}
  	} 
  	\quad 
  	\subfigure[]{
    \includegraphics[width=0.9\columnwidth]{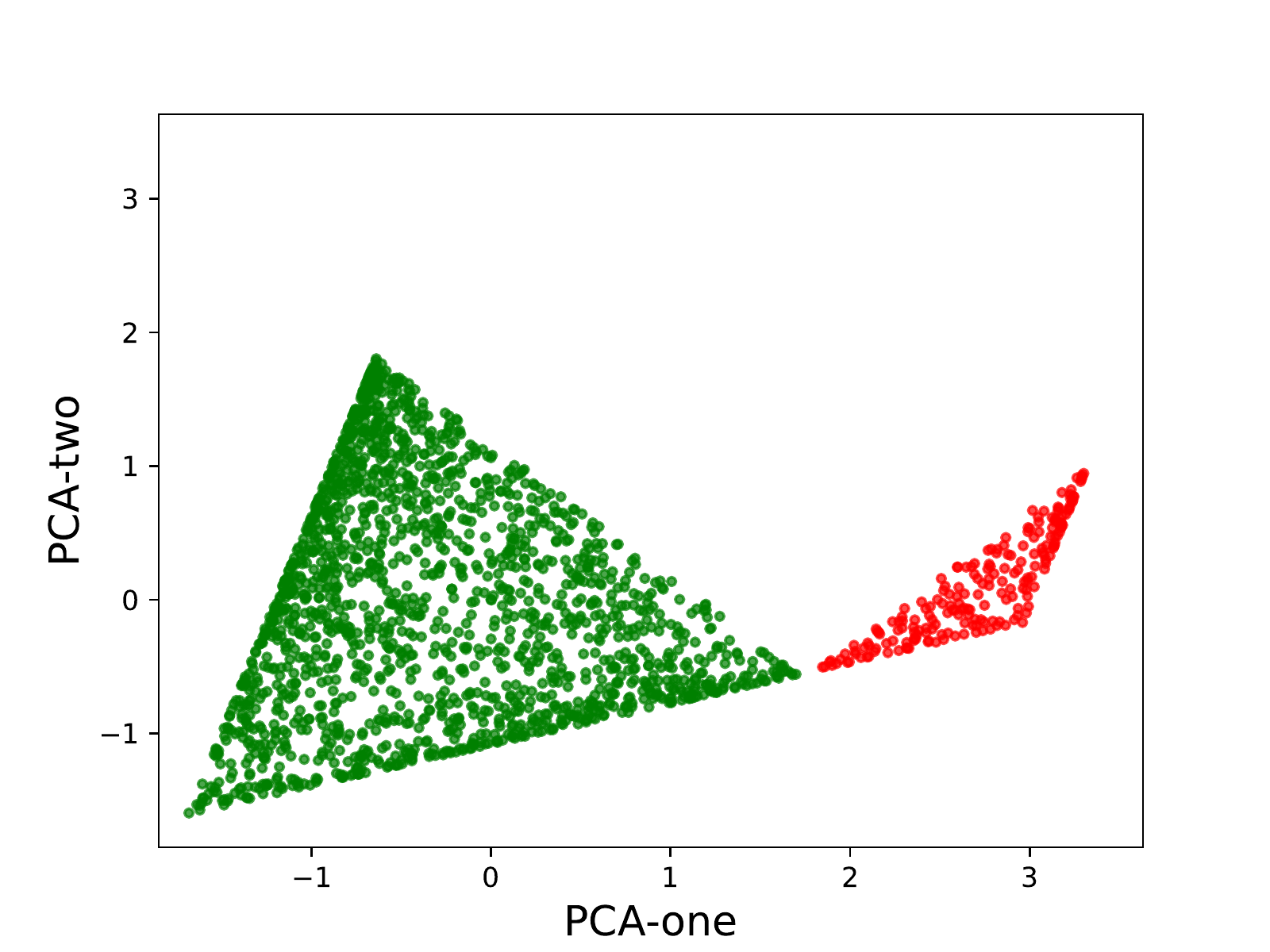} 
	\label{fig:2_wx_b}  	
  	} 
  	\caption{Communities detected in the benchmark network. (a) Original network, showing communities 1 and 2 colored in, respectively, green and red. (b) PCA of the node distance vectors.}
	\label{fig:2_wx}
\end{figure*}

A network with two communities was created using the model described in Section~\ref{s:benchmark}. The parameters used in the model were $R_1=3$, $R_2=1$ and $\lambda_1=\lambda_2=65$. The distance between the two reference points for each community was set to $R_1+R_2$. The created network was partitioned by the proposed method assuming Euclidean distance between the nodes. The weight parameter was set to $\alpha_{c_i}=R_i$. The result is shown in Figure~\ref{fig:2_wx_a}. The method correctly identified all the nodes in the two communities.

In order to better visualize the association of the nodes with each community, Principal Component Analysis (PCA) was applied to the node distance vectors. The result is shown in Figure~\ref{fig:2_wx_b}. Two well-separated clusters are observed in the PCA space. 

\begin{figure}
	\begin{center}
    	\includegraphics[width=0.9\columnwidth]{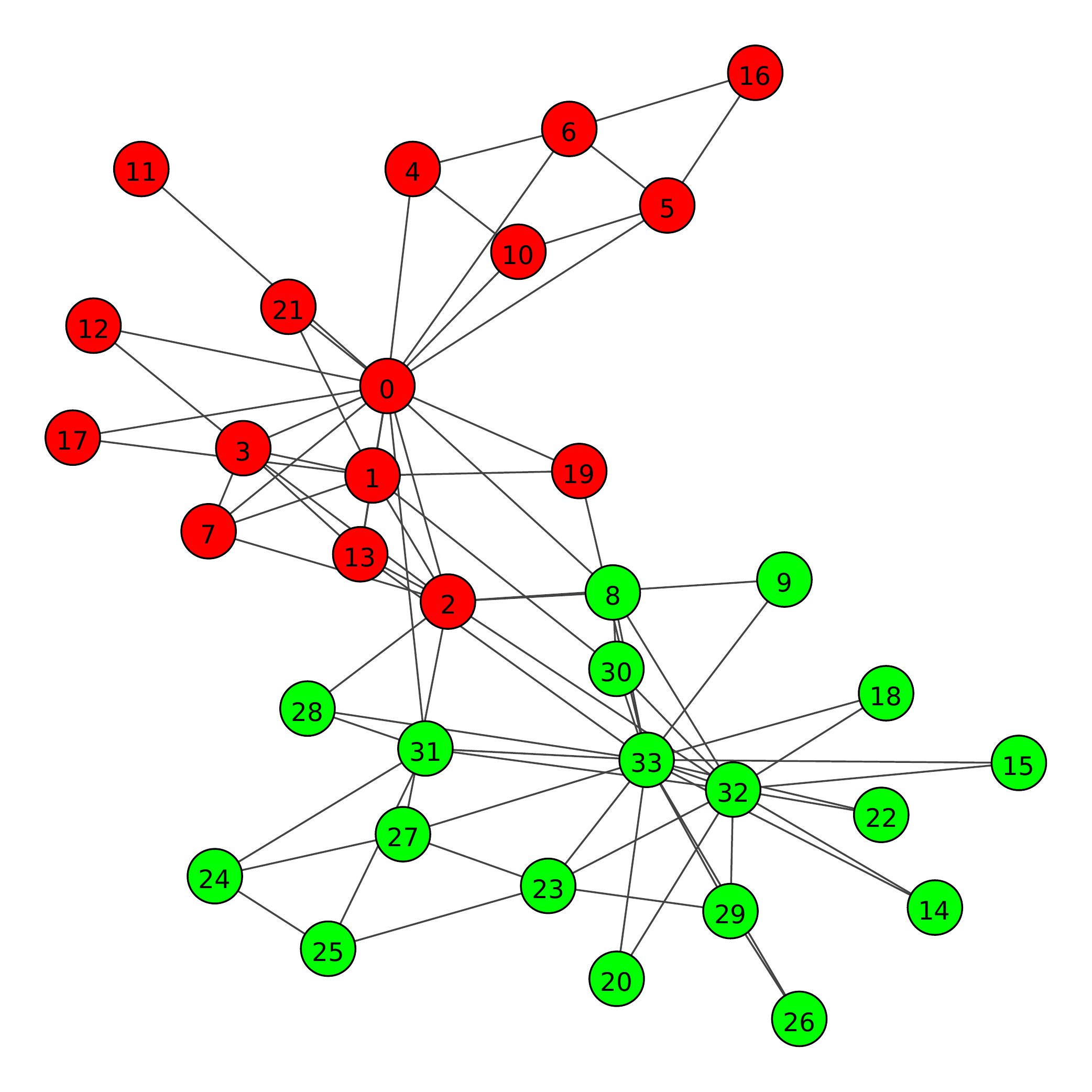}
    \end{center}
    \caption{Detected communities in the Zachary's karate club.}
    \label{fig:kara}
\end{figure}

The proposed method was also applied to the Zachary's karate club network~\cite{zachary1977information}. The two nodes with the largest degree, $v_{i=0}$ and $v_{i=33}$, were set as reference for each community. The topological distance, instead of a geometric distance, was used and the weights were set to $\alpha_{c_1}=\alpha_{c_2}=1$. The detected communities are shown in Figure~\ref{fig:kara}.

\section{Concluding Remarks}

Community finding remains a challenging problem in network science.  In this note, we propose a distance-based method that takes into account the topological or geometric distances to pre-defined reference points.  The membership of each node is determined by taking the community identifier corresponding to the smallest of the obtained distances for each node.  The method has been found to perform surprisingly well for several spatial networks.  Future developments should include the evaluation of the methodology for other spatial networks, as well as its extension to non-spatial structures.      

\acknowledgments
Paulo Paulino thanks CNPq-USP (grant no. 157326/2017-9) for support. C. H. Comin thanks FAPESP (grant no. 15/18942-8) for financial support. L. da F. Costa thanks CNPq (grant no. 307333/2013-2) and NAP-PRP-USP for sponsorship. This work has been supported also by FAPESP grants 11/50761-2 and 2015/22308-2.

\bibliographystyle{apsrev}
\bibliography{references1}
\end{document}